\documentstyle[12pt]{article}
\advance\textheight by \topskip
\begin{document}
\input epsf

\begin{flushright}
CERN-TH/96-66\\
SU-ITP-96-10\\
hep-th/9603090\\
March 13, 1996\\
\end{flushright}

\begin{center}
\baselineskip=16pt

{\Large\bf      UNIVERSALITY ~OF
\vskip .5cm
 SUPERSYMMETRIC ~ATTRACTORS}  \\

\vskip 1.8cm
 {\bf Sergio Ferrara}\footnote{E-mail:
ferraras@cernvm.cern.ch}\\
 \vskip 0.2cm
Theory Division, CERN, 1211 Geneva 23, Switzerland\\
\vskip .6cm

{\bf Renata Kallosh}\footnote {E-mail:
kallosh@physics.stanford.edu}\\
 \vskip 0.2cm
Physics Department, Stanford University, Stanford   CA 94305-4060, USA\\
\vskip .6cm

\end{center}

\vskip 1 cm
\centerline{\bf ABSTRACT}
\vspace{-0.3cm}
\begin{quotation}

The macroscopic entropy-area formula for supersymmetric black holes in  N=2,4,8
 theories  is found to be universal:  in d=4 it is  always given by the square
of the largest of the central charges  extremized in the moduli space. The
proof of universality is based on the fact that the doubling of unbroken
supersymmetry near the black hole horizon  requires that all central charges
other than Z=M vanish at the attractor point for N=4,8.  The ADM mass at the
extremum can be computed in terms of  duality symmetric quartic invariants
which are moduli independent.
The extension of these results for d=5, N=1,2,4 is also reported.
A duality symmetric expression for the energy  of the ground state  with
spontaneous breaking of  supersymmetry is  provided by  the power 1/2 (2/3)  of
the black hole area of the horizon in d=4 (d=5).\\
\vspace{-0.3cm}

It is suggested that the  universal duality symmetric formula for the
energy of the ground state in supersymmetric gravity is given by the modulus of
the maximal central charge at the attractor point in any supersymmetric theory
in any dimension.

\end{quotation}

\normalsize
\newpage

\normalsize

\section{Introduction}

The  attractor behavior in the supersymmetric system was discovered in the
context of N=2  extremal magnetic black holes \cite{FKS}.  It was explained in
this work that
unbroken supersymmetry leads to the fact that the area of the black hole
horizon depends only on conserved charges.
The idea was further developed for more general  black hole solutions in
\cite{S,FK}. In   \cite{FK} the complete
treatment of d=4 N=2 and d=5 N=1 supersymmetric attractors  was presented as
well as
 some particular examples of N=4,8 attractors in d=4. The main result of this
work was that
the area of the supersymmetric black hole horizon in the theories of N=2
supergravity interacting with arbitrary number of vector and hypermultiplets
can be computed by extremizing the central charge of the theory in the moduli
space.
The summary of the attractor picture in N=2 theories is the following. There is
only one  central charge $Z_{AB}$ where $A,B= 1,2$ and the central charge
$Z_{AB}$ is complex and antisymmetric in $A,B$. Therefore we have only
$Z_{12}=Z$.
Unbroken supersymmetry  of the N=2 black holes requires
\begin{equation}
M^2_{\rm ADM} (q,p,\phi) = |Z(q,p,\phi)|^2 \ .
\end{equation}

In gravitational theories the ADM mass in asymptotically flat spaces defines
the energy of the space-time  $M_{\rm ADM}=E$.
Thus in N=2 theories one may have concluded that the area of the horizon is
proportional to the square of the energy in its minimum.
\begin{equation}
{A(q,p) \over 4} = \pi  E^2 (q,p,\phi)| _{{\partial E \over \partial \phi}=0} \
{}.
\label{area1}\end{equation}
Vice versa,  the  minimal energy is given by the square root of the area
\begin{equation}
E (q,p)=\left  ({A(q,p) \over 4\pi  }\right )^{1/2}  \ .
\label{area2}\end{equation}
In d=5 the minimal energy in N=2 theories was shown to be $E(q,p) \sim  \left
(A(q,p) \right )^{2/3}$ \cite{FK}.

The  purpose of this paper is  to find out how much of this behavior carries
over to higher extended supersymmetries, where the number of central charge
eigenvalues is $N/2>1$.  We would like to mention in this respect that the
focus of interest to the  problem as to whether the area always does
not depend on moduli  was  stimulated by the work of  Larsen and Wilczek
\cite{LW}.

The area formulas for  theories with extended supersymmetries are build out of
central charges on the basis of duality invariance and comparison with some
known particular black hole solutions.
For example, in N=4 theories  there are two eigenvalues of the central charge
matrix $Z_1$
and $Z_2$.  The area of the horizon for the  particular black holes in N=4
supergravity without matter multiplets is proportional
to  \cite{US}
\begin{equation}
\left ({A(q,p) \over 4}\right )_{N=4} = \pi (|Z_1| -|Z_2|^2) \ .
\label{N4}\end{equation}
In N=8 theory the corresponding formula for the area depends on 4 eigenvalues
of the central charge matrix and is given in agreement with some class of
solutions by \cite{KallE(7)}
\begin{equation}
\left ({A(q,p) \over 4}\right )_{N=8}= \pi \left( {  |(\sum\limits_i |Z_i|^4 -2
\sum\limits_{i>j} |Z_i|^2 |Z_j|^2 + 8
|Z_1 Z_2 Z_3 Z_4| })| \right)^{1/2}\ ,
\label{N8}\end{equation}
Formula
(\ref{N8}) corresponds to the Cremmer-Julia \cite{CJ}  E(7) quartic invariant
\begin{equation}
\diamondsuit =  {\mbox Tr} \Bigl(Z\bar Z \Bigr)^2 -{\textstyle{1\over 4}}
\Bigl({\mbox Tr}\,Z\bar Z\Bigr)^2 + 4  \Bigl({\mbox P\hskip- .1cm f}\; Z +
{\mbox P\hskip- .1cm f} \; \bar Z \,\Bigr)  \
{}
\label{diamond}\end{equation}
in the normal frame for the central charge matrix \cite{FSZ}.
Each central charge depends on moduli and electro-magnetic charges.

One may have thought  that the function
to be extremized  to get the area will be some combination of various central
charges in $N>2$  theories generalizing the one central charge situation in N=2
theories. However, this is not the case, the result of our study is: the
unbroken supersymmetry leads to the {\it universal  formula for the area of the
horizon} in all d=4 extended supersymmetric theories. First, let us note that
the energy of the space-time, or   ADM mass is equal to the largest eigenvalue
of the central charge matrix   from the requirement of unbroken
supersymmetry\footnote{ For simplicity we consider only even N.}.
\begin{equation}
E(q,p,\phi )=M_{\rm ADM} (q,p,\phi)=   max \,  |Z_C (q,p,\phi)|   \ , \qquad
C=1, \dots, N/2.
\end{equation}
We will find that the square of the minimal energy always defines the area of
the horizon of the supersymmetric black holes.
\begin{equation}
{A(q,p) \over 4} = \pi  E^2 (q,p,\phi)\vert _{\phantom{}_{{\partial E \over
\partial \phi}=0} } \ .
\label{area3}\end{equation}
Let us relabel the central charges: the largest one $ max \,  |Z_C  (q,p,\phi)|
$will be called   $|Z|$  so that
$$ max \,  |Z_C  (q,p,\phi)|  \equiv |Z|\ , $$
and the remaining $ ({N\over  2}-1)$ eigenvalues of the central charge matrix
will be labelled by the index  $c$ which runs from  1 to  $({N\over  2}-1)$.

We will establish that the fixed point of attraction in the theories of
extended supersymmetries with $N>2$ is
given by the condition of the vanishing of all eigenvalues of the central
charge matrix which are smaller than the largest one, defining the ADM mass.
\begin{equation}
 |(Z_c )_{\rm fix} | =0\ ,  \qquad c = 1, \dots ,({N\over  2}-1).
\end{equation}
The area formula therefore is always given by the extremum value of the central
charge, or space-time energy, which is  the point where the other charges
vanish.
Let for example in N=4 case the first eigenvalue is larger than the second
$|Z_1| >|Z_2| $.
We will find that the area formula for black holes in N=4 supergravity with N=4
vector   multiplets is given by a duality symmetric formula which is also an
extremum of the
ADM mass defined by the  vanishing of the next to the largest eigenvalue of the
central charge.

$A(q,p)$ can be explicitly computed, by noticing that $\pi (|Z_1|
-|Z_2|^2)_{Z_c=0}=\pi (|Z_1| -|Z_2|^2) $ at the matter attractor point
(deduced by setting the gaugino variation $\delta \lambda^{ia}=0$ but not
fixing the S-dilaton).
In N=4 we have that $|Z_2|=0$ forces the dilaton to take the fixed value
$S=S_{\rm fix}$ and  the matter  scalars to the point  $\delta
\lambda^{a}_i(a=1,\dots , n_v)=0$. So
\begin{equation}
(|Z |^2)|_{\rm fix}= (|Z_1| -|Z_2|^2)_{S=S_{\rm fix}, \, \delta
\lambda^{a}_{i}=0} \ .
\end{equation}
But now we use the fact that  $ (|Z_1| -|Z_2|^2)_{ \delta \lambda^{a}_{i}=0}
$ is independent of S and this allows to give an explicit formula for
$|Z_1|_{\rm fix}$ in terms of 2(6+ $n_v$) charges.
\begin{equation}
|Z_1|^2_{\rm fix}= {1\over 2} \sqrt {q^2 p^2 - (q\cdot p)^2} \ .
\end{equation}
Similarly in  N=8, the attractor point is the value of the 70 moduli $\phi =
\phi _{\rm fix}$ for which $|Z_2|= |Z_3|=|Z_4|=0$. This means that the area
formula is given in terms of the   Cartan's quartic invariant J which
depends only on charges and does not depend on moduli. The universality of the
area formula comes from the fact that the Cremmer-Julia E(7) invariant at the
attractor point one   hand has to depend only on charges and  be E(7)
symmetric, and therefore has to coincide with Cartan's  invariant J, on the
other hand at the attractor it is given by
\begin{equation}
\left ({A(q,p) \over 4}\right )_{N=8}= \pi \left( {  |(\sum\limits_i |Z_i|^4 -2
\sum\limits_{i>j} |Z_i|^2 |Z_j|^2 + 8
|Z_1 Z_2 Z_3 Z_4| })| \right)^{1/2}_{|Z_2|= |Z_3|=|Z_4|=0} = |Z_1|^2_{\rm fix}\
,
\label{N82}\end{equation}
Therefore
\begin{equation}
|Z_1|^2_{\rm fix} =( \sqrt {\diamondsuit}) _{\phi = \phi _{\rm fix}} =
\sqrt {J} \ ,
\end{equation}
and this gives an explicit formula of $|Z_1|^2_{\rm fix}$ in terms of the 56
charges.

If Cremmer-Julia E(7)   invariant $\diamondsuit $ is $\phi$ independent in the
generic point of the moduli space,
 which seems likely, in this case  we have also
\begin{equation}
 \diamondsuit = J  \ ,
\end{equation}
as conjectured by Cremmer and Julia \cite{CJ}. In any case  at the attractor
point these two invariants coincide and this is the reason for the universality
of our area formula.

Similar results are also obtained for N=2,4 at d=5.

We  find it useful to introduce here an additional object $\Sigma(\rho)^I$ for
any scalar field $\phi^I (r)$, which forms part of the black hole solution. It
is  typical for the attractor problem to have a pair of phase space variables
$(x(t), y(t) = x'(t))$.  In our case the corresponding pair consists of   the
scalar field  and the first derivative of the scalar field
\begin{equation}
\Sigma^I (\rho) \equiv   {\partial  \over \partial \rho} \phi^I (\rho) \ .
\end{equation}
We have plotted the value of the dilaton $e^{-2\phi(\rho)}$ and the $\Sigma
(\rho)$ for the dilaton $U(1)^2$ black hole. We are using here the radial
variable  $\rho = -1/r$ in terms of which the near horizon geometry is
conformally flat. Fig. 1a) shows that starting with any initial conditions at $
\rho =  0$  ($r     \rightarrow  \infty$) the field is driven to an attractor
value at $ \rho    \rightarrow - \infty$  ($r=0$).
Fig. 1b) shows how the derivative evolves. The initial condition for  the
derivative of the scalar field at $ \rho = 0$  ($r     \rightarrow  \infty$) is
equal to the so-called scalar charge $\Sigma$ . It is a function of the moduli
at $ \rho = 0$  $ (r \rightarrow  \infty) $ and electromagnetic charges,
therefore the scalar charge is also called a secondary hair of the black hole.
When evolving into the core of the black hole, independently of the initial
value of the scalar charge at  $ \rho = 0$  ($r     \rightarrow  \infty$), the
derivative (i.e. the effective scalar charge)   $\Sigma(\rho)$ goes to zero  at $ \rho =  \rightarrow - \infty$ $ (r=0)$.

The physical picture of this phenomenon was suggested to us by A. Linde. It
reflects the fact that the values of electric and magnetic charges are
protected by gauge invariance and the associated Gauss low. One can measure the
actual values of electric and magnetic  charges being far away from the black
hole, which explains why these charges are called the black hole hair. The
scalar charges which can be measured far away from the black hole are not
protected by any conservation low. This is why the derivative of the scalar
fields  decreases when moving into the core of the black hole, see Fig. 2b).
Whereas the electric and magnetic charges are conserved and their presence and
stability supports the existence of an infinite throat of the Bertotti-Robinson
\cite{BR} geometry, the derivative of the scalar charge $\Sigma(\rho)
\rightarrow 0$ at $\rho \rightarrow \infty$, i.e. it does  not
penetrate into the throat to keep its size (the area of the black hole horizon)
minimal. This results in the minimal energy of the ground state. This picture
may apply   not only for supersymmetric black holes. One may try
to find out if the analogous phenomenon takes place at least for near extreme
black holes, as suggested to us by L. Susskind. The plausibility of this
picture relies on the fact that we may be dealing here with the critical
phenomena which have some specific range of applicability and in particular the
attractor behavior of the system may well describe the nearby trajectories in
the phase space of the system.

One should stress that the extremization of the energy  which we study here is
performed under condition that the scalar fields are constant as it is usually
done in quantum field theory by looking for the minima of the energy in the
class of configurations with constant scalars.
 This is the main feature of supersymmetric attractors, since we study the
fixed points of the differential equations where the scalars have vanishing
derivatives \cite{FKS}.
\begin{equation}
\phi' _{\rm fix} (\rho) \rightarrow 0 \ , \qquad \rho \rightarrow   \infty \ .
\end{equation}
The  massless black holes (solutions with  ADM mass going to zero)    which one
would tend to associate with the minimum of the ADM energy, do not fit  into
the class of attractors with constant scalars in the fixed point and has to be
studied separately. 

From the point of view of supersymmetry, this distinction comes from the following  fact. The graviphoton charge, which is a linear combination of moduli and electro-magnetic charges, always  represents the ADM mass for supersymmetric solutions, i.e. the energy of the space-time. The minimization in the class of configurations with constant scalars by supersymmetry requires all vector fields of the matter multiplets
to vanish at the attractor. Their charges also are given by some linear combinations of moduli and electro-magnetic charges, which vanish at attractor.
This leads to Bertotti-Robinson-type geometry and black holes with non-vanishing area of the horizon. 

On the other hand one may be interested in the opposite situation when the graviphoton charge tends to zero, leading to a specific relation between the moduli and electro-magnetic charges, corresponding to  massless black holes. The matter multiplet   charges do not vanish, otherwise the solution would be trivial. By supersymmetry it follows that the scalars do not tend to a constant near the black hole core, the geometry is very different from the 
Bertotti-Robinson-type geometry and this class of configurations has to be studied separately.

Also the configurations with the zero area which have  doubled or quadrupled number of unbroken supersymmetries comparatively to those with non-vanishing area, do not  have fixed points for the scalar fields near the horizon \cite{KP}.

In what follows we will explain how the result described  above follows from
unbroken supersymmetry.  In Sec. 2 we will explain the particular case of pure
N=4 supergravity, the $U(1)^2$ model. We will  present the complete theory of
attractors
in  N=4 supergravity interacting with arbitrary number of N=4 vector
multiplets in
Sec. 3. The area formula is presented in  various forms which show the
manifest  S and T duality as well as moduli independence as the consequence of
supersymmetry. On top of it the area formula is finally reduced to the
universal one in terms of the minimization of the    ADM mass in the
moduli space.
 In Sec. 4. we will analyze the  simple N=8 model, an STU attractor, which
presents in a nice way the gross features of the attractors with extended
supersymmetries.      The   general N=8  attractor is described  in Sec. 5.
Cartan's quartic invariant which reflects the E(7) symmetry of the theory is
miraculously reduced to the simple and universal formula for the area as the
minimum of the largest eigenvalue of the central charge.  Sec. 6 presents
supersymmetric attractors in  d=5. We present the area formulas in  N=1,2,4
theories.
In particular, the cubic E(6) invariant is described in connection with 3
central charges of the maximally extended N=4 theory in d=5. The  universality
of the area formula again follows from
the vanishing of all eigenvalues of the central charges except the largest one.
One more miracle of supersymmetry and we get
the universal formula for the area in terms of the minimization of the 
ADM energy  for all  cases considered.

These results should apply to strings theories or generalization thereof
(M-theory?).
A microscopic derivation is expected to correct the entropy formula for small
values of charges. This is of course related to the fact that Einstein gravity
is only the point-like limit approximation of the more general theories.

Our  study of supersymmetric black holes and their attractor behavior suggest
the following interpretation: we have computed  the exact   energy of the
ground state in supersymmetric gravity as the function of electric and magnetic
charges. The  presence of these charges  (excitations of a superstrings) in the
vacuum leads to spontaneous breaking of supersymmetry, resulting in  a duality
symmetric positive energy of the ground state associated with the non-vanishing
area of the black hole horizon.   The recent success in the calculation of the
area of the horizon of some 5 and 4   dimensional black holes \cite{Microsc}
from the point of view of counting string states with the use of  D-brane
technology \cite{Pol}   naturally fits into our interpretation. The non-trivial
part of this picture is related to the  fact that the extreme black holes with
the non-vanishing area of the horizon have a conformal isometry  which
exchanges the two asymptotic regions.
Therefore the  calculation of the area of the horizon of such extreme black
holes can be interpreted as the evaluation  of the square of the energy of the
ground state of this system. The corresponding conformal factor has one
parameter,  the area of the horizon, or the mass of the Bertotti-Robinson
universe.
The relevance of the computation of the area of the extreme black hole horizon
to spontaneous supersymmetry breaking  is explained in Sec. 7 of the paper.

\section{Attractor in pure N=4 supergravity, $U(1)^2$ model}
The basic feature of the extended supersymmetry attractors can be easily
understood already in the case of pure N=4 supergravity with one gravitational
multiplet only. The bosonic fields in the $SU(4)$ version  consist of a complex
axion-dilaton scalar, three vectors and three axial vectors and the metric. It
was explained in \cite{KP} that near the horizon the unbroken supersymmetry of
the $U(1)^2$ black hole is doubled. Instead of 1/4 of N=4 supersymmetry, near
the horizon 1/2 of N=4 supersymmetry is restored. In the basis chosen in
\cite{US}
we have the following situation near the horizon:
 for positive $pq$ the unbroken $N=2$ supersymmetry consists of the
third
and the fourth ones, and for negative values of $pq$ it is reverse,
the first
and
the second supersymmetries are unbroken whereas the third and the
fourth are
broken.
The
dilatino transformations rules in the notation of
\cite{US} is,
\begin{equation}
\frac{1}{2}\delta \Lambda_I = - \gamma^\mu
\epsilon_I \partial_\mu\phi + \frac{1}{\sqrt 2} \sigma^{\mu\nu}\left(
e^{-\phi}
F_{\mu\nu} \alpha_{IJ} -
e^\phi \tilde G_{\mu\nu}\beta_{IJ}\right)^-\epsilon^J = 0\ .
\label{dilatino}\end{equation}
The first term in (\ref{dilatino}) vanishes at the fixed point, since we are
looking for
$\phi'=0$.
Thus at the attractor we get   for $pq>0, \;  J=3,4$ as well as for $pq<0, \;
J=1,2$
the second term in
eq.
(\ref{dilatino})

\begin{equation}\label{controller}
\Sigma_{\rm fix } \,  \epsilon^J = {\textstyle{1\over 2}}({\mbox{e}}^{-\phi}
|p| -  {\mbox{e}}^{\phi}  |q|)_{\rm fix } \,  \epsilon^J= 0,
\end{equation}
which leads to the condition of the vanishing of the dilaton charge at the
attractor:
\begin{equation}
\Sigma_{\rm fix }   = {\textstyle{1\over 2}}({\mbox{e}}^{-\phi}
|p| -  {\mbox{e}}^{\phi}  |q|)_{\rm fix } \,  = 0  \qquad  \Longleftrightarrow
   \qquad {\mbox{e}}^{-2\phi}_{\rm fix }= {|q|\over |p|} \ .
\end{equation}

We may also rewrite the dilatino transformation rule at the attractor  in the
form
\begin{equation}\label{controller2}
(Z_{IJ})_{\rm fix }  \epsilon^J = 0.
\end{equation}
For $pq>0, \;  \epsilon^3, \epsilon^4$ are non-vanishing, therefore at the
attractor
using also the  gravitino transformation rule we  learn that
\begin{equation}
Z_{34}=0 \ , \qquad   |Z_{12}| =(M_{ADM})_{Z_{34}=0} \qquad pq>0 \ ,
\end{equation}
and for  $pq<0, \;  \epsilon^1, \epsilon^2$ are non-vanishing, therefore at the
attractor
\begin{equation}
Z_{12}=0 \ , \qquad   |Z_{34}| =(M_{ADM})_{Z_{12}=0} \qquad pq<0 \ .
\end{equation}

\section{Attractor in N=4 supergravity with $n_v$ vector multiplets: general
case}
The geometry of the N=4 supergravity coupled to $n_v$ matter vector multiplets
\footnote{We describe here the version of N=4 theory closely related to the one
in \cite{Eric}. The version here has the property of being symplectic
covariant. The details of this construction will be presented elsewhere.}
 is based on the non-linear sigma model ${SU(1,1)\over U(1)} \times
{O(6,n_v)\over  O(6) \times O(n_v)}$. The ${SU(1,1)\over U(1)}$ manifold is
parametrized by a complex scalar field $S$ and the vector multiplet manifold by
the coset representatives $L_\Lambda ^A = (L_\Lambda ^{ij}, L_\Lambda ^a)$:
\begin{equation}
L_\Lambda ^{ij} = - L_\Lambda ^{ij} = L_{\Lambda ij}^* = {1\over 2} \epsilon
^{ijkl} L_{\Lambda kl} \ , \end{equation}
where $ i,j = 1,2,3,4, \quad \Lambda = 1, \dots , 6+n_v, \quad a= 1, \dots,
n_v
$,
and there are  orthogonality relations
\begin{equation}
-L_\Lambda^a L_{a\Sigma} + L_\Lambda ^{ij} L_{\Sigma ij} = \eta_{\lambda
\Sigma}\ ,
\qquad L_\Lambda^a L^\Lambda_b =- \delta^a_b \ ,
\end{equation}
and
\begin{equation}
L_\Lambda ^{ij} L^\Lambda _{kl} = {1\over 2} (\delta^i_{[k} \delta ^j_{l]}) \ ,
\qquad
L_\Lambda^a L^\Lambda _{ij} =0 \ .
\end{equation}
The vector field (complexified) kinetic matrix is
\begin{equation}
{\cal N} _{\Lambda \Sigma} = (S- \bar S ) L_\Lambda ^{ij} L_{\Sigma ij} + \bar
S\,  \eta _{\Lambda \Sigma} \ ,
\end{equation}
and the symplectic sections are
\begin{eqnarray}
( L^\Lambda _{ij} , \; {\cal N} _{\Lambda \Sigma} L^{\Sigma}_{ij} &=& S\,
L_{\Lambda ij} ) \ , \nonumber\\
 \nonumber\\
( L^\Lambda _{a} , \; {\cal N} _{\Lambda \Sigma} L^{\Sigma}_{a} &=& \bar S\,
L_{\Lambda a} ) \ .
\end{eqnarray}
In terms of these sections the central charge $Z_{ij}$ is
\begin{equation}
Z_{ij} = e^{K/2} [ L^\Lambda_{ij} q_\Lambda - S L_{ij \Lambda} p^\Lambda ] \ ,
\end{equation}
where $K = - \log i (S-\bar S)$ is the $S$-field Kahler potential.
For the $x$-independent scalars $(S, \, L^\Lambda _{ij} , \, L^\Lambda _{a} )$
 unbroken supersymmetry for the matter gaugino's $\delta \lambda_i^a =0$
requires at the attractor point
\begin{equation}
S L_\Lambda^a p^\Lambda -  L^{a\Lambda} q_\Lambda =0\ ,
\label{N=4 attr}\end{equation}
while unbroken N=1 supersymmetry for the dilatino $\chi^i$ requires that  the
central charge $|Z_2| < |Z_1| = M_{ADM}$  given by
\begin{equation}
|Z_2|^2 ={1\over 4}  \left (Z_{ij} \bar Z^{ij} -  \sqrt { (Z_{ij} \bar Z^{ij} )^2 -
{1\over 4} |\epsilon^{ijkl}  Z_{ij} Z_{kl} )|^2 }\right) 
\label{S}\end{equation}
should vanish. Equation (\ref{S}) fixes the value of S at its attractor point:
\begin{equation}
Z_2=0 \ .
\end{equation}
It can be proved using the symplectic formulation of N=4 theory given above
that the quantity
\begin{equation}
|Z_1|^2 - |Z_2|^2= {1\over 2} \sqrt { (Z_{ij} \bar Z^{ij} )^2 - {1\over 4}
|\epsilon^{ijkl}  Z_{ij} Z_{kl} )|^2 }
\end{equation}
is $S$-independent. This would be  sufficient  to prove the moduli independence
of this expression in pure N=4 supergravity. However, in presence of N=4 vector
multiplets this expression as a function of attractor variables corresponding
to $\alpha, \beta $ charge vectors of string theory does depend on the scalars
of these multiplets (the asymptotic value of the matrix ${\cal M}$ used in
various black holes constructions). This was established in \cite{DLR} (see eq.
(8.13) of this paper) starting with ten dimensional supersymmetry and using the
Witten-Israel-Nestor construction.  However our formalism shows that at the
matter attractor point defined in eq. (\ref{N=4 attr}) this expression does not
depend on matter moduli anymore and becomes the function of charges only:
\begin{equation}
(|Z_1|^2 - |Z_2|^2)_{|Z_2| =0}  = {1\over 2} \sqrt {q^2 p^2 - (q  \cdot p)^2 } \ ,
\end{equation}
where Lorentzian  $O(6, n_v)$ norm for $q_\Lambda, \, p^\Lambda$ doublet is
understood.
It then follows that the area is
\begin{equation}
A=2\pi \sqrt {q^2 p^2 - (q  \cdot p)^2 } \ .
\label{heter}\end{equation}
Note that the 2(6+$n_v$) electric and magnetic charges form an $SU(1,1)$
doublet and (6, $n_v$) Lorentzian vectors.
This coincides with the minimum of the ADM mass
\begin{equation}
A(q,p) = 4 \pi (M^2_{ADM} )_{{\partial M \over \partial \phi}=0}
\end{equation}
in the axion-dilaton  and  matter moduli space.

We would like to stress here that in \cite{FK} we have introduced the concept
of attractor variables for the black hole solutions: variables in which the ADM
mass depends on charges and moduli however,  the area depends only on charges.
 The area  formula (\ref{heter}) in the attractor variables  appeared in the
recently revised version of ref. \cite{CT}.   Our choice of what are  attractor
variables is defined by the formulation  of the theory with manifest symplectic
symmetry. Simultaneously this form  provides the proof of the independence of
the area from the moduli
 and leads to the universality of the area formula in terms
of the extrema of the ADM mass in the moduli space.
On the other hand the heterotic area formula given in \cite{CT} as well as our
proof of independence on all moduli at the attractor point applied to the
expression for central charges in \cite{DLR} provides the link to the
properties of a string theory.  Indeed  the corresponding attractor variables
are the conserved $\alpha, \beta$ charge vectors of string theory introduced
into the black hole physics by Sen \cite{Sen} long time before it was realized
that the area  of supersymmetric black holes depends only on
 $\alpha, \beta$ by  the reason of supersymmetry, as explained in this paper.
Our N=4 results and formulas, since they entirely rely on general theory of N=4
supergravity coupled to vector multiplets, should equally apply to the
heterotic string compactified on $T_6$ or to the type II string compactified on
$K_3\times T_2$.

 \section{ STU Attractor  in N=8 $SU(8)$ Supergravity}

In \cite{FK} we have described     STU model   in the attractor variables.
Here, as the preparation to general N=8 attractor we would like to check
whether the main principle of the minimization of the largest eigenvalue of the
central charge  will produce the moduli independent area.
We will denote $
e^{-\eta_0} = {\rm Im} S=s, \quad e^{-\sigma _0} = {\rm Im} T=t,  \quad
e^{-\rho_0} = {\rm Im} U =u.$
The ADM mass considered as a function of charges  in generic point of the
moduli space $(s,t,u)$ is
\begin{equation}
M_{ADM} = {1\over 4} \Bigl(stu|q_1|  +{s \over tu}|q_3| +
{u \over st }|p_2| +{t \over su}|p_4|\Bigr) \ .
\end{equation}
The variation of the mass over the moduli gives 3 attractor equations:
\begin{eqnarray}
&&\Bigl(stu|q_1| +{s \over tu}|q_3| -
{u \over st }|p_2|  -{t \over su}|p_4|\Bigr)=0\ ,\nonumber\\
\nonumber\\
&&\Bigl(stu|q_1| -{s \over tu}|q_3|+
{u \over st }|p_2|  -{t \over su}|p_4|\Bigr)=0\ ,\nonumber\\
\nonumber\\
&&
\Bigl(stu|q_1| -{s \over tu}|q_3|-
{u \over st }|p_2|  +{t \over su}|p_4|\Bigr)=0 \ .
\label{attractor}\end{eqnarray}
The solution of these equations puts the moduli into the fixed points where
they become functions of charges.
\begin{equation}
(stu)_{\rm fix}\, |q_1|= \left ({u \over st }\right )_{\rm fix}|p_2|=\left ({s
\over tu}\right )_{\rm fix}|q_3|=\left ({t \over su}\right )_{\rm fix}|p_4| \ .
\end{equation}
We get the useful relations
\begin{equation}
(s^2 t^2)_{\rm fix}= {|p_2|\over |q_1|} \ , \qquad (u^2 t^2)_{\rm fix}=
{|q_3|\over |q_1|} \ , \qquad (s^2 u^2)_{\rm fix}= {|p_4|\over |q_1|} \ ,
\end{equation}
and
\begin{equation}
(stu)_{\rm fix} = \left( {|p_2 q_3 p_4| \over |q_1|^3}\right)^{1/4}  .
\end{equation}
This allows to get the value of the ADM mass at the attractor:
\begin{equation}
(M_{ADM})_{\rm fix} = (stu)_{\rm fix}|q_1|  = |q_1p_2 q_3 p_4|^{1/4} \ .
\end{equation}
We may now conjecture that the generalization of the 4 eigenvalues of the N=8
supergravity central charges \cite{KallE(7)}  is the following
\begin{eqnarray}
&&4 Z_1=  \Bigl(stu|q_1|  +{s \over tu}|q_3|\Bigr) +
\Bigl({u \over st }|p_2| +{t \over su}|p_4|\Bigr) \ ,
\nonumber\\
\nonumber\\
&& 4 Z_2=  \Bigl(stu|q_1|  +{s \over tu}|q_3|\Bigr) -
\Bigl({u \over st }|p_2| +{t \over su}|p_4|\Bigr)\ ,\nonumber\\
\nonumber\\
&& 4 Z_3=  \Bigl(stu|q_1|  -{s \over tu}|q_3|\Bigr) +
\Bigl({u \over st }|p_2| -{t \over su}|p_4|\Bigr)\ ,\nonumber\\
\nonumber\\
&& 4 Z_4=  \Bigl(stu|q_1|  -{s \over tu}|q_3|\Bigr) -
\Bigl({u \over st }|p_2| -{t \over su}|p_4|\Bigr)\ .
\end{eqnarray}
Now one can see that indeed the 3 attractor equations (\ref{attractor}) mean
exactly the following:
\begin{equation}
Z_2=Z_3=Z_4=0 \ , \qquad Z_1|_{\phantom{}_{Z_2=Z_3=Z_4=0}}=(M_{ADM})_{\rm fix}
=  |q_1p_2 q_3 p_4| =E(p,q)\ .
\end{equation}
This example makes it natural to look for the general N=8 attractor expecting
to get the area from the minimum in the moduli space of the largest of the 4
eigenvalues of the central charge matrix.

\section{ N=8  attractor: general case}

N=8 theory has only one gravitational multiplet. Therefore all 28 vector fields
are graviphotons, there are no vector fields which are not supersymmetric
partners of the graviton.  The 28 electric and 28 magnetic charges all together
are in 56 fundamental representation of $E_7$. The black hole solutions of this
theory with 1/8 of supersymmetry unbroken are known to have a non-vanishing
area. The manifestly E(7) symmetric area formula is given by the unique quartic
invariant of E(7). However, how can we find out if the area is independent on
70 moduli and depends only on 56 charges?
Let us first analyze the supersymmetry transformation  rules \cite{CJ}  near
the attractor where all 70 moduli tend to a constant values.
\begin{eqnarray}
\delta \Psi_{\mu A} &=&  D_\mu \epsilon _A+  Z_{AB\, \mu\nu} \gamma^\nu
\epsilon^B \ ,\\
\nonumber\\
\delta \chi _{ ABC } &=&    Z_{[AB  \, \mu\nu} \sigma ^{\mu \nu}  \epsilon_{C]}
 \ .
\label{susy}\end{eqnarray}
Let us decompose N=8 into $SU(4)\times SU(4)$ as (4,1) + (1,4), keeping all
fields of N=8 theory. Then N=8 supergravity multiplet will split into one
gravitational multiplet of N=4 theory
$$[ (2)    \quad 4(3/2) \quad 6(1) \quad 4(1/2) \quad  2(0)]$$
4 spin 3/2 multiplets
$$4 [  (3/2) \quad 4(1) \quad 6+1 (1/2) \quad  8(0)]$$
and 6 vector multiplets
$$6 [  (1) \quad 4 (1/2) \quad  6(0)].$$
The 8-dimensional index $A$ is split as $A= (i,a)$ where $i=1,\dots ,4$  and
$a=1,\dots ,4$. The fermions are $ \Psi_{\mu A} = (\Psi_{\mu i} \ , \Psi_{\mu
a}) $ and
$ \chi _{ ABC } = (  \chi _{ ijk }\ ,  \chi _{ iab }\ ,  \chi _{ aij }\ ,  \chi
_{ abc })$. In N=4 theory $ \chi _{ ijk }$ is in spin 2 multiplet, $\chi _{ aij
}\ ,  \chi _{ abc }$ belong to spin 3/2 multiplet and $ \chi _{ iab }$ to spin
1 multiplet. We may solve eqs. (\ref{susy})
using the ansatz
\begin{equation}\label{f1}
\epsilon_a = 0, ~~~\epsilon_i =\{\epsilon_1,  \epsilon_2 \not= 0,~~~\epsilon_3
= \epsilon_4= 0 \}   .
\end{equation}
The transformation of 4 gravitino from the gravitational multiplets and of
those from the spin 3/2 multiplets are
\begin{eqnarray}\label{f2}
\delta\psi_{\mu i} &=& D_\mu\epsilon_i + Z_{ij\, \mu\nu}\gamma^\nu\epsilon^i\ ,
  \\
\nonumber\\
\delta\psi_{\mu a} &=& D_\mu\epsilon_a + Z_{ai \, \mu\nu}\gamma^\nu\epsilon^i\
{}.
\end{eqnarray}
The central charge matrix can be put into the  normal frame \cite{FSZ} by
means of  a SU(N) transformation. In this frame the off diagonal elements
$Z_{ai} $ are absent,  $Z_{ai}=0 $.
\begin{equation}
Z_{AB} = \pmatrix{
Z_{ij}  & 0 \cr
0 & Z_{ab} \cr
}.
\end{equation}
 Matrices  $Z_{ij}$ and $Z_{ab}$ are diagonal,
\begin{equation}\label{f22}
Z_{ij} = \pmatrix{
z_1\sigma_2 & 0 \cr
0 &z_2\sigma_2 \cr } \ , ~~~~~~~Z_{ab} = \pmatrix{
z_3\sigma_2 & 0 \cr  0 & z_4\sigma_2 \cr
} \ ,
\end{equation}
where
\begin{equation}\label{f4}
\sigma_2 = i\pmatrix{
0 & 1 \cr
-1 & 0 \cr
}.
\end{equation}
Thus the transformation of 4 gravitino from spin 3/2 multiplets vanishes
$\delta\psi_{\mu a} =0$ due to the fact that 4 $\epsilon_a$ supersymmetries are
broken and that in the normal frame the off diagonal elements of the central
charge matrix $Z_{ai} $ are absent.
The variation of spin 1/2 fields vanishes
\begin{equation}\label{f5}
\delta\chi_{abc} = 0, ~~~\delta \chi_{aij} = Z_{ij}\epsilon_a +
Z_{ia}\epsilon_j = 0 \ .
\end{equation}
It remains to   check the spin 1/2 transformations from the gravitational
multiplet and from the vector multiplet of N=4 theory at the attractor point :
\begin{equation}\label{f6}
\delta\chi_{iab} = Z_{ab\mu\nu}\sigma^{\mu\nu}\epsilon_i \ ,
\end{equation}
\begin{equation}\label{f7}
\delta\chi_{ijk} = Z_{ij\mu\nu}\sigma^{\mu\nu}\epsilon^j \ .
\end{equation}
The first of these two equations implies that $Z_{ab} = 0$, i.e. $z_3 = z_4 =
0$. The second one yields $z_2 = 0$, as in $N = 4$. Thus we have proved that
the attractor condition indeed requires all eigenvalues of the central charge
which are smaller than the one equal to the ADM mass to vanish. This again
leads us to the universal formula for the area, starting with an E(7) symmetric
formula.

 \section{Attractors  in d=5 N=2,4 theories}
The general formulas for N=4, N=8 derived in this paper have an obvious
extension at d=5. In our previous paper \cite{FK} we gave the general area
formula for an arbitrary N=1 theory in d=5 in terms of the symmetric constant
$d_{ABC}$
\begin{equation}
A\sim Z^{3/2}_{\rm fix} = \left (d^{AB}(q)^{-1} q_A q_B \right)^{3/4} \ ,
\end{equation}
where $d^{AB}(q)^{-1}$ is the inverse of the moduli dependent matrix $d_{AB}=
d_{ABC} t^C$
computed at the attraction point $Z=Z_{\rm fix}$. There was only one central
charge $Z= t^A
(z)q_A$  ( $d_{ABC} t^A t^B t^C =1$) and $\partial_i Z=0 \Longrightarrow
Z=Z_{\rm fix}$

For N=2, N=4 in d=5 we have 2 and 3 central charges respectively\footnote{The
reason why in N=4 supergravity in d=5 there are only 3 central charges in the
normal frame is due to the fact that the $Z_{ij}$  (i,j=1,\dots 8) central
charge matrix is traceless
$
Z_{ij} \Omega^{ij} =0
$
with respect to the $Sp(8)$ metric $\Omega^{ij} = - \Omega^{ji}$. By reducing
d=5 to d=4 on $S_1$ one gets a fourth charge from Kaluza-Klein vector $g_{\mu
5}$. }
 \cite{5d}. Again the ADM mass is given by the largest eigenvalue. Let us call
it $Z$ and the other $Z_c$. As before
\begin{equation}
A\sim (Z)^{3/2}_{\rm fix} \ .
\end{equation}
For N=2 coupled to $n_v$ matter  multiplets there are $6+ n_v$ electric
charges. They are in the
(5,$n_v$) vector representation of O(5, $n_v$) + a singlet. The singlet is the
charge of the vector dual to the $B_{\mu\nu}$ field.

The general formula for $Z$ at the attractor point coincides with the macroscopic formula given by Strominger and Vafa \cite{Microsc},
\begin{equation}\label{s1}
Z|_{\phantom{}_{\rm fix}} = (Q_HQ^2_F)^{1/3}, 
\end{equation}
where $Q_H$ is the singlet charge and $Q_F^2$ is a Lorentzian $(5,  n_\nu)$ norm of the other $5 + n_\nu$ charges.

For the $N = 4$ theory we have 27 charges which are in the 27 irreducible representation of $E_6$. The formula for $Z$ is  given by the cubic root of the unique $E_6$ invariant constructed out of the 27 dimensional representation of $E_6$, which is the central charge $Z_{ij}$ $(i,j = 1, ..., 8)$. (Note that the 27  can be represented as a traceless $Sp(8)$ symplectic matrix.)
\begin{equation}\label{s2}
Z|_{\phantom{}_{\rm fix}}= (\triangle)^{1/3} = \Bigl( q_{ij} \Omega^{jl}q_{lm} \Omega^{mn} q_{n\rho}\Omega^{\rho i}\Bigr)^{1/3} \ , 
\end{equation}
where $q_{ij}$ is a 27 integer charge vector transforming under $E_6({\bf Z})$ (integer valued $E_6$ group).

These $N = 4$ results are expected to apply to the type II strings compactified on the $5$-torus or eleven-dimensional supergravity (or M-theory) on the six-torus.

\section{Exact Total-Energy of the Ground State and Spontaneous Breaking of
Supersymmetry}

We would like to
use here the  experience  from the study of  supersymmetric black holes in
string theory, accumulated in the community over the recent years and also some
ideas from duality symmetric quantization of superstring theory \cite{Quant} to
learn about the properties of the exact Hamiltonian in quantum theories with
local supersymmetry. It has been pointed out in \cite{Quant} that the
consistent quantization of $\kappa$-symmetry in the backgrounds with unbroken
supersymmetry can be performed with the help of the supercharge of the
background in  which the extended $\kappa$-symmetric object can exist.
The supercharge of the gravitational supersymmetric theory was defined
by Teitelboim \cite{Supercharge}   as the surface
integral
in terms of the gravitino $\Psi_{\mu}$ field of the configuration, solving the
field equations:
\vskip -.3cm
\begin{equation}
{\cal Q} = \oint_{\partial \Sigma}  d\Sigma_{\mu\nu} \gamma^{\mu\nu\lambda}
\Psi_\lambda  \ .
\end{equation}
\vskip -.3cm
The surface over which the integration has to be performed depends on the
choice of configuration. In all cases it is the same surface the integration
over
which defines the ADM mass of a given system or the  ADM mass per unit  area
(length).
 The on-shell backgrounds with some number  of supersymmetries
unbroken in bosonic sectors have the vanishing supersymmetry variation of the
gravitino, when the parameters are Killing spinors.\vskip -.3cm
\begin{equation}
{\cal Q}_{k}  = \oint_{\partial \Sigma}  d\Sigma_{\mu\nu}
\gamma^{\mu\nu\lambda}
 \delta_{\epsilon_k} \Psi_\mu  =
 \oint_{\partial \Sigma}   d\Sigma_{\mu\nu}
\gamma^{\mu\nu\lambda}
\hat \nabla _\lambda \epsilon_k =0 \ .
\label{killingcharge}\end{equation}
 For anti-Killing spinors the supercharge is not vanishing. For the black hole
multiplets it defines the
so-called superhair of the black hole:
\vskip -.3cm
\begin{equation}
{\cal S}_{\rm superhair} \equiv {\cal Q}_{\bar k}  =  \oint_{\partial \Sigma}
d\Sigma_{\mu\nu}
\gamma^{\mu\nu\lambda}
\hat \nabla _\lambda \epsilon_{\bar k}  \ .
\end{equation}
\vskip -.3cm
\noindent The concept  of the {\it superhair}  was defined for the first time
for
 extreme Reissner-Nordstr\"om  black holes in  \cite{Aich} and studied more
recently in the context of more general extreme black holes in \cite{BrKO}.

In the theories with local supersymmetry the total-energy operator
(Hamiltonian) \cite{DT} is defined via the quadratic combination of
supercharges
\begin{equation}
P_0 = (8\pi Gh N)^{-1} \sum_{I=1}^N \, \sum_{A=1}^4 {\cal Q}_{AI}^2~.
\end{equation}
The computation of the area of the horizon of the supersymmetric  black holes
performed above via the extremum of the ADM energy suggest the following
interpretation:  the ADM mass at the extremum in the moduli space  is the value
of the
total-energy operator (Hamiltonian) of the ground state. The ground state has a
non-vanishing vacuum energy due to the presence of  electric and magnetic
charges, which cause spontaneous breaking of supersymmetry.
\begin{equation}
H_{vac} = E(p,q) = \sqrt {A(p,q)\over 4\pi }~.
\end{equation}
The electric and magnetic charges are due to the excitation of the
microscopical
degrees of freedom of the string theory or alternatively due to the near
horizon black hole geometry of Bertotti-Robinson, which exists only when the
charges are non-vanishing. For the string theory interpretation of the ground
state energy the appropriate formulas are:
 for the heterotic string compactified on $T_6$ (or type II compactified on
$K_3 \times T_2$)  the $SL(2,{\bf Z}) \times
SO(6,22; {\bf Z})$ symmetric expression for the ground state energy is
\begin{equation}
E(p,q)_{\rm het}  = \sqrt {A(p,q)\over 4\pi } = \left (q^2 p^2 - (q \cdot
p)^2\right )^{1/4},
\label{hetenergy}\end{equation}
where the 2 (6+22) electric and magnetic charges form $SL(2,{\bf Z})$ doublets
and (6,22) Lorentzian vectors.
For Type II string compactified on $T_6$ the vacuum energy is given by the
$E_7({\bf Z})$ symmetric expression.
\begin{equation}
E(q,p)_{\rm {II}} = J^{1/4} \ ,
\label{II}\end{equation}
where the quartic Cartan $E(7)$ invariant is
\begin{eqnarray}
J &=&  q^{ij}  p_{jk}  q^{kl}  p_{li} - {\textstyle{1\over 4}} q^{ij}  p_{ij}
q^{kl}  p_{kl} \nonumber\\
 \nonumber\\
&+& {\textstyle{1\over 96 }}   \Bigl(\epsilon^{ijklmnop} \,  p_{ij}  p_{kl}
p_{mn}  p_{op}  + \epsilon_{ijklmnop} \,   q^{ij}   q^{kl}   q^{mn}   q^{op}
\Bigr) \ .
\label{Car}\end{eqnarray}
 and the charges $q^{ij }$ and  $ p_{ij}$, $i,j = 1, \dots , 8$ span the  56
dimensional space.
Our interpretation is supported by the following space-time picture.
The black hole configurations interpolate between two vacua, one trivial at
asymptotic infinity and the second one
 described by the Bertotti-Robinson geometry, which is known to have an
unbroken N=2 supersymmetry \cite{Gibb,K,KP,FK}. In particular, due to conformal
flatness of the geometry and due to covariantly constant
graviphoton field strength, the supersymmetry variation of the gravitino field
strength vanishes without enforcing any linear combination of the supersymmetry
parameter to vanish.
\begin{eqnarray}
&&\delta_{\rm susy}  \;  (D_{[\mu} \psi_{\nu]})_{BR} =0 , \qquad \epsilon^1
\neq 0, \qquad \epsilon^2 \neq 0, \nonumber\\
&&\delta_{\rm susy}  \;  (D_{[\mu} \psi_{\nu]})_{triv} =0 , \qquad \epsilon^1
\neq 0, \qquad \epsilon^2 \neq 0 .
\end{eqnarray}
At asymptotic infinity the trivially flat vacuum is also characterized by the
unbroken space-time supersymmetry. Moreover, for the trivial vacuum the
variation of the gravitino itself vanishes, since both spin connections as well
as vector field strengths vanish at asymptotic infinity.
Consider now the second vacuum, the near horizon configuration of the
supersymmetric black holes. The space is only conformally flat, it is
characterized by
some electric and magnetic charges. The unique parameter, characterizing the
geometry, the Bertotti-Robinson mass, is given by duality symmetric function of
all available electric and magnetic charges. One can check that despite the
supersymmetry transformation of the field strength of gravitino in this
background vanishes the transformation of gravitino does not for non-vanishing
charges $p,=q,p$. This causes the main difference with the trivial
asymptotically flat vacuum with zero energy.  \begin{eqnarray}
&&\delta_{\rm susy}  \;  ( \psi_{\nu})_{BR} \neq 0 , \qquad \epsilon^1 \neq 0,
\qquad \epsilon^2 \neq 0,\\\label{spont}
&&\delta_{\rm susy}  \;  ( \psi_{\nu})_{triv} = 0 , \qquad \epsilon^1 \neq 0,
\qquad \epsilon^2 \neq 0 .
\end{eqnarray}
The ground state  energy, which is proportional to the square root of the area
of the black hole horizon (or to the Bertotti-Robinson mass) is positive,
duality symmetric
and presents a non-trivial computation of an eigenvalue of the energy-operator
of  a ground state  of quantum gravity system.
The energy of the ground state
\begin{equation}
(P_0)_{vac} = (8\pi Gh N)^{-1} \sum_{I=1}^N \, \sum_{A=1}^4 ({\cal
Q}_{AI})_{vac}^2
\end{equation}
is always non-negative, however, for the ground state to have a positive energy
one has to require that the supercharge of the ground state is non-vanishing.
This happens in our system since the ground state supercharge
does not vanish in presence of the covariantly constant graviphoton field
strength:
This makes the  calculation of the ground state energy of the theory with local
supersymmetry consistent with the idea of spontaneous breaking of supersymmetry
with  the non-vanishing constant value of the vacuum supercharge.

\section{Discussion}

In this paper we have established the universality of the black hole area
formula for  extended supersymmetric theories $N\geq 2$ in d=4. It is based on
the fact that all central charges of the theory except the one which equals the
ADM mass have to vanish near the black hole horizon by requirement of
supersymmetry. In this way supersymmetry realizes the principle of lowest
possible ground state. The fact that the ADM mass as a function of charges and
moduli is equal to the central charge does not mean yet that it is a ground
state. Since the central charge depends on conserved electric and magnetic
charges and moduli in the generic point the energy is not the minimal one. One
has to minimize it in the moduli space and this is how we get the minimal
energy of the ground state.

This universality can be  understood also  from the fact that in supersymmetric
theories with asymptotically flat spaces there exists a well defined universal
expression for the Hamiltonian in terms of the sum over all supercharges
\cite{DT}.
The presence of an extended microscopic object like superstring introduces
spontaneous breaking of the supersymmetry, from the point of view of the
space-time Hamiltonian, since it has  a non-vanishing value on the ground state
of the system.
The electric and magnetic charges which are interpreted as charges $(q,p)$
defining the  Bertotti-Robinson geometry and defining the size of its infinite
throat $A(q,p)$,  from the point of view of string theory are simply the
conserved charge vectors of string theory$(\vec \alpha, \vec \beta)$.

If one accepts the point of view that the calculation of the area of the black
hole horizon was a tool to get the ground state energy, it become clear that
any supersymmetric black hole solution with particular area formula actually
provides the calculation of the ground state energy and gives a specific
example of the ground state energy calculation. However, the general expression
for the energy of the ground state is simultaneously universal and duality
invariant as explained in this paper,
\begin{equation}
E(q,p) = E(\vec \alpha, \vec \beta) = \left ({A(q, p)\over 4 \pi }\right)^{1/2}
\ .
\end{equation}
This result can be associated with the fact discovered by Gaillard and Zumino
\cite{GZ} that the energy momentum tensor in supergravities is duality
invariant whereas the off-shell Lagrangian is not.
Our analysis also predicts that the most general  S and T duality invariant
area formulas  in the heterotic theory compactified on $T_6$ (or type II,
compactified on $K_3\times T_2$ , given in eq.  (\ref{hetenergy})  or with  U
duality for type II string theory compactified on $T_6$ as   given in eqs.
(\ref {II})  should be reachable by the counting of string states, as it was
already demonstrated in particular examples \cite{Microsc}.

The space-time picture is  that the most  general 4 dimensional supersymmetric
black holes with the non-vanishing area of the horizon, covering the
singularities,   interpolate between 4 dimensional Minkowski space-time ${\bf
M}^4$, at spatial infinity $r\rightarrow \infty$ and $adS_2\times S^2$ down the
infinite wormhole throat $r\rightarrow 0$ as was noticed by Gibbons \cite{Gibb}
 with respect to Reissner-Nordstr\"om geometry. In addition the near horizon
geometry upon the change  of variables
\begin{equation}
r= - {E^2(p,q) \over \rho}
\end{equation}
was also interpreted in \cite{FK} as the second conformally flat  ${\bf M}^4$.
The conformal factor
relating these two asymptotic regions  ${\bf M}^4(r) $  and ${\bf M}^4(\rho ) $
was  found  to be equal to  \cite{FK}
\begin{equation}
{E(q,p)^2 \over \rho^2} = {r^2 \over E(q,p)^2} \ .
\end{equation}
Thus the total stability of this picture of the space-time geometry is
constrained severely by the fact that the space-time energy of the ground state
$E(q,p)$ is not vanishing. The existence  of these charges is explained by the
existence of the string states. Moreover, from the point of view of string
theory the two  ${\bf M}^4$ do not seem to be distinguishable. This is one
possible explanation of the spontaneous supersymmetry breaking behind the
non-vanishing of the ground state energy
$E(q,p)\neq 0$.
When the energy of the ground state vanishes and the area of the black hole
horizon shrinks to zero this picture of two asymptotically  ${\bf M}^4 $
regions is not valid anymore, scalar fields do not stop evolving inside the
throat. In the duality symmetric area formulas  the vertices described in
\cite{US,KallE(7)} are reached, double or quadruple number of supersymmetries
is restored and singularities become naked unless the ground state energy
$E(q,p)$ is non-vanishing.
Our macroscopic formulae
are supposed to be valid for large values of charges but one may
expect corrections to them from microscopic physics, for example one can find some   disagreement between microscopic and macroscopic calculation of the entropy for small charges \cite{Microsc}.  Therefore the status of   
the zero entropy limit  may be  changed by microscopic physics.

\section*{Acknowledgements}
We have had most fruitful and enlightening discussions of the results of this
work with   E. Bergshoeff,  A. Linde, T. Ort\'{\i}n, L. Susskind, and E. Verlinde,  and an
interesting correspondence with E. Cremmer  and B. Julia on E(7) invariants.

S.F. was  supported in part
by DOE under grant DE-FGO3-91ER40662, Task C and by EEC Science program
SC1*ct92-0789 and INFN.
 R.K. was  supported
by  NSF grant PHY-9219345.

\vfill
\newpage

\begin{figure}
\centerline{ \epsfxsize 4in\epsfbox{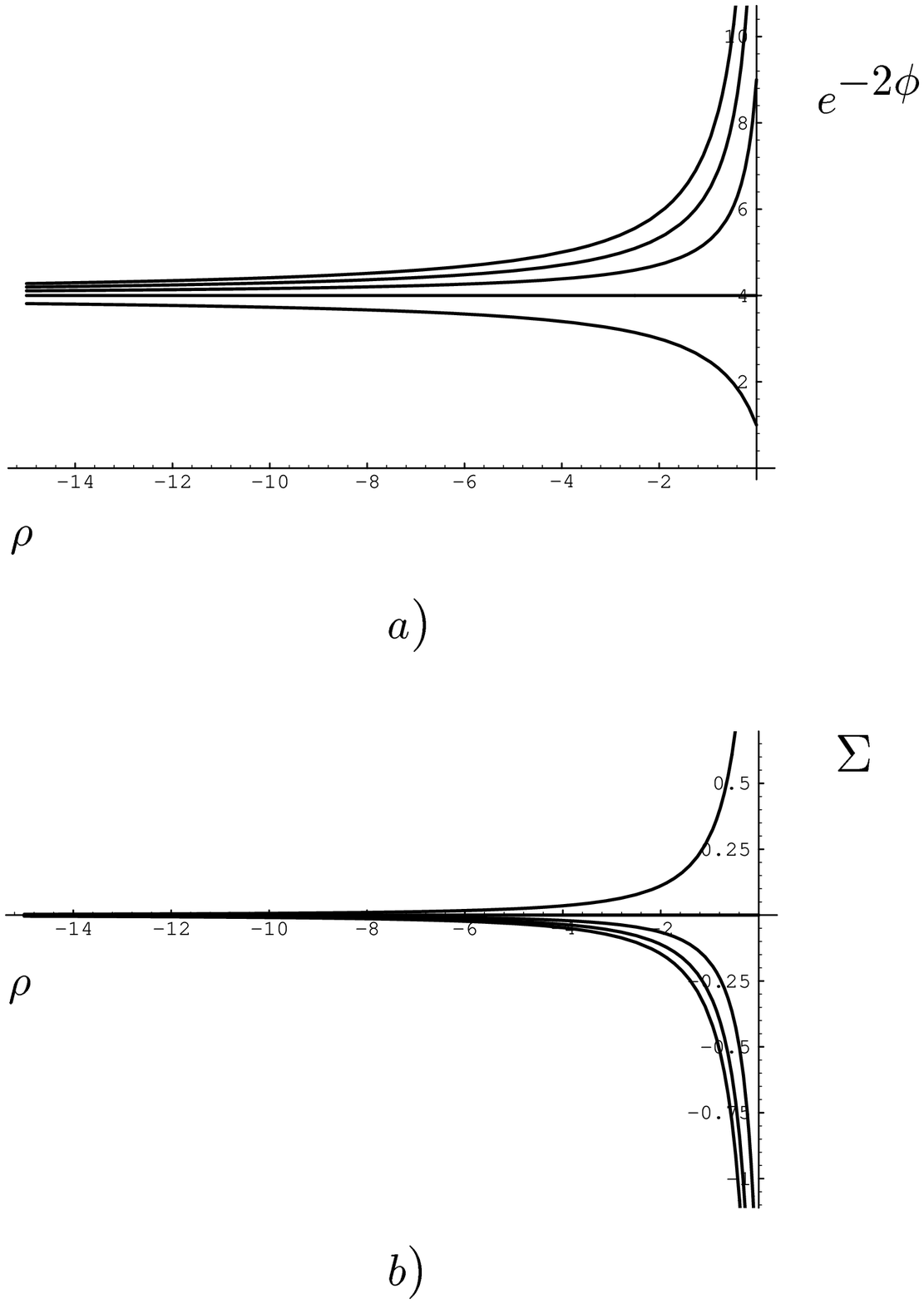}}
 \vskip 1.5cm
\caption{Evolution of the dilaton field $e^{-2\phi}$ and of the effective dilaton charge $\Sigma(\rho) = - {d\phi\over d\rho}$ for various initial conditions at $\rho = 0$ ($r = \infty$) to a common fixed point
at $\rho =-\infty$ ($r = 0)$.}

\label{F22}

\end{figure}

\end{document}